# *Micrometer-Thin Crystalline-Silicon Solar Cells Integrating Numerically Optimized 2-D Photonic Crystals*

*V. Depauw[1*], X. Meng[3,6], O. El Daif[1], G. Gomard[3,7], L. Lalouat[3,4,5], E. Drouard[3,4], C. Trompoukis[1,2], A. Fave[3,5], C. Seassal[3,4], I. Gordon[1]*

[1] Imec, Kapeldreef 75, 3001 Leuven, Belgium

[2] Katholieke Universiteit Leuven, Departement Elektrotechniek – ESAT, Kasteelpark Arenberg 10, B-3001 Leuven, Belgium

[3] Université de Lyon, Institut des Nanotechnologies de Lyon (INL) UMR 5270 CNRS- INSA-ECL-UCBL, France

[4] Ecole Centrale de Lyon, 36 Avenue Guy de Collongue, 69134, Ecully, France

[5] INSA de Lyon, Bat. Blaise Pascal, 7 Avenue Capelle, 69621, Villeurbanne, France

[6] Presently at McMaster University, 1280 Main Street West, L8S 4L7, Hamilton, Ontario, Canada

[7] Presently at Light Technology Institute (LTI), Karlsruhe Institute of Technology (KIT), 76131 Karlsruhe, Germany

*\* corresponding author, valerie.depauw@imec.be*

.

## *Abstract*

A 2-D photonic crystal was integrated experimentally into a thin-film crystalline-silicon solar cell of 1-µm thickness, after numerical optimization maximizing light absorption in the active material. The photonic crystal boosted the short-circuit current of the cell, but it also damaged its open-circuit voltage and fill factor, which led to an overall decrease in performances. Comparisons between modeled and actual optical behaviors of the cell, and between ideal and actual morphologies, show the global robustness of the nanostructure to experimental deviations, but its particular sensitivity to the conformality of the top coatings and the spread in pattern dimensions, which should not be neglected in the optical model. As for the electrical behavior, the measured internal quantum efficiency shows the strong parasitic absorptions from the transparent conductive oxide and from the back-reflector, as well as the negative impact of the nanopattern on surface passivation. Our exemplifying case, thus, illustrates and experimentally confirms two recommendations for future integration of surface nanostructures for light trapping purposes: 1) the necessity to optimize absorption not for the total stack but for the single active material, and 2) the necessity to avoid damage to the active material by pattern etching.

# I.   Introduction

The photovoltaic market is currently dominated by the so-called first generation technology, with which crystalline silicon (c-Si) wafers of about 180-µm thickness are processed into solar cells and assembled into a module. The second generation, based on a few micrometers, or less, of amorphous silicon or of more exotic materials (CdTe, CuInGaSe2, etc.), deposited on large panels, have been gaining shares thanks to their lower processing costs. However, contrary to what was prophesized in the early days, and as their name indicated, they have not overtaken c-Si cells yet. Research for first generation cells has in fact been able to incrementally and continuously push their efficiency upwards. Unexpectedly high levels have been reached today, with as recent example, Schott Solar reaching 19.9% with the mainstream industrial Al-back surface field cells concept [1]. This efficiency increase has been coupled with a decrease in wafer thickness, since one of the main cost drivers of this technology is the amount of silicon consumed per Wattpeak [2]. However, this cost reduction process is now hitting a wall [3], as wafers are becoming too thin, and thus, fragile for cell manufacturers. Besides, module assembly costs have become another major cost component. An obvious breakthrough would, therefore, be to switch from a cell-to-module process to an integrated cell-and-module process [4], [5], with which thin foils of c-Si would be transferred to a glass panel as early as possible in the process. This would simultaneously solve handling issues, by providing support to the foils, and decrease the module assembly costs, by combining cell and module steps and by simplifying cells interconnection. This hybrid technology would combine the advantages of the first generation (efficiency, stability, established technology, etc.) with those of the second (simpler fabrication, low material consumption).

A thin-film approach for c-Si solar cells has, however, a major challenge, that is the low absorptivity of c-Si for the red part of the sunlight. To keep a high energy-conversion efficiency, thin-film c-Si cells require advanced light trapping concepts, not only to specifically trap red photons, but also because standard light management, based on micrometer-size pyramid texturing, is not suited for micrometer-thin silicon foils. Advanced light trapping is nowadays a very active field, and various photonics concepts can be found in recent literature, as reviewed by Mokkapati and Catchpole [6]. Plasmonics [7], front and back gratings [8], [9], nanowires [10], [11], random textures [12], or combinations [13], [14] are being investigated for various cell technologies, including thin c-Si cells. For the latter, the works presented are, however, usually discussing optical demonstrators, and yet only few c-Si solar cells have been shown [15]–[18]. Therefore, if the impact of these concepts for photon collection is well studied, their impact on charge-carriers collection is much less.

This study addresses integration in a thin c-Si solar cell of a 2-D photonic crystal (2DPC). Indeed, among the previously mentioned concepts, the 2DPC have been shown numerically and experimentally to be a promising approach for increasing light absorption in the active thin c-Si layer, which provides a good control of the optical properties and avoids adding absorbing materials [8], [19], [20]. We present here the next step of a full cell demonstrator, hence taking into account not only optical [21] but also electrical aspects. The solar cell is fabricated on an ultrathin film of 1 µm and the optically optimal nanostructure is patterned at its front side. The device is, then, characterized topographically, optically, and electrically to identify the various integration challenges that nanogratings bring for charge collection. Given its extreme thinness and its simple design (with limited surface passivation), the maximum reachable efficiency of such solar cell is limited, but it remains an ideal reference case for light trapping and passivation studies.

## II. *Experimental setup*

Detachable monocrystalline silicon films were transferred to the glass, nanopatterned by laser holographic lithography and dry etching, and then, processed into heterojunction solar cells. The material fabrication and cell processing followed the methods of "epifree" cells described in [22]. To integrate the surface nanopattern, we decided to pattern the c-Si film prior to forming the junction, in order to avoid the need for top-contact alignment and to facilitate surface passivation. For reference purposes, a sample was processed in parallel but without nanopatterning. The processes are chronologically and briefly described below (more extensive descriptions can be found in [21]–[23]).

The targeted geometrical parameters of the 2DPC have been previously optimized thanks to finite-difference time domain (FDTD) calculations on a slightly simplified structure and using the typical optical indices of the various involved materials to maximize the sunlight absorption in the active layer [21].

### A. *Material preparation and rear-side cell processing*

The monocrystalline micrometer-thin film was prepared by annealing a patterned wafer at high-temperature, and then, processed to form a back-surface field (BSF) and a base contact before transferring to a glass substrate. The material, 1.2-µm thick, was formed using the empty-space-in-silicon technique, by which cylindrical well-ordered pores, etched by deep-UV lithography and dry etching, are reorganized and merged into a single buried void by annealing above 1000 °C in a non-oxidizing atmosphere [see Fig. 1(a)]. During this thermal treatment, a 250-nm thick layer of boron-doped silicon was epitaxially grown from dichlorosilane at 900 °C to form a BSF. This method, which is not compatible with a cost-effective thin-film process, has however, the advantage here of letting us integrate some rear-side surface passivation, which is necessary for the purpose of observing the impact of the front-side pattern, without unnecessary risks to harm the fragile suspended film in extra steps. The BSF was $10^{19}$ at/cm$^3$ doped, while the base was $10^{15}$ at/cm$^3$. On top of the p++ epitaxial layer, the base contact was formed by depositing 1-μm-thick film of aluminum by e-beam evaporation. This aluminum side was bonded anodically to a glass substrate under 1000 V at 250 °C for 10 min. The film, then, directly detached from its parent wafer.

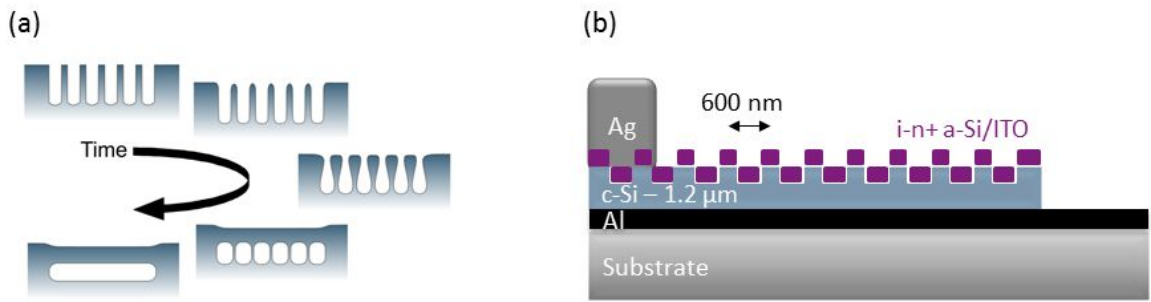

**Fig. 1. (a) Empty-Space-in-Silicon technique by which the c-Si thin film forms at 1130 ºC, and (b) Epifree a-Si:H heterojunction solar cell of 1-μm-thick monocrystalline silicon integrating a photonic crystal.**

## B. *Nanopatterning*

The nanopatterning method was extensively described in [21]. The c-Si film stacked on Al and glass was first covered with a SiO$_2$ hard mask by plasma-enhanced chemical vapor deposition (PECVD). The laser-interference lithography, which is a versatile technique well suited for large area patterning, is then used to pattern a chemically assisted photoresist. The pattern is then transferred into the oxide, and then, the silicon layer using, respectively, reactive ion etching and inductive-coupled plasma etching. The whole process enables a sufficient verticality of the side walls of the nanopattern. Finally, the remaining silica mask is removed using buffer HF and O$_2$ plasma.

### C. *Front-side cell processing*

Subsequent to the nanopatterning step, the hydrogenated amorphous silicon (a-Si:H) emitter was deposited and contacted. After wet cleaning, a stack of intrinsic and n+ hydrogenated amorphous silicon, 25-nm thick in total, was deposited by PECVD at 170 °C and covered by a nominal thickness of 75 nm of sputtered indium tin oxide (ITO) for lateral conductivity and antireflection effect. Fingers of Ti/Pd/Ag were evaporated through shadow masks and annealed at 200 °C in $N_2$ to improve their adhesion. Finally, 1 cm × 1 cm solar cells [see Fig. 1(b)] were defined and separated by lithography and reactive ion etching ($SF_6/O_2$ plasma).

### D. *Characterization*

The samples were characterized topographically, optically and electrically. The pattern profile and dimensions were investigated by atomic force microscopy (AFM), scanning electron microscopy (SEM, secondary electrons detector) and transmission electron microscopy (TEM). For TEM, the samples were capped by CVD $SiO_2$ and lifted out by focused ion beam milling. The device reflectance (R) was measured using a spectrophotometer on the finished cells in between two metal fingers, within a wavelength range from 300 to 1100 nm and integrated over the whole half space, using an integrating sphere. The spot size was 1 mm × 1 mm. The spectral step was 10 nm and the spectral width was ∼2 nm. Absorption of the full stack is then given by A = 1–R–T, with T = 0 from the presence of the aluminum rear contact. For angle-resolved measurements, a different spectrophotometer was used, a Perkin Elmer Lambda 750S. The measurements were performed in an integrating sphere of 150-mm diameter with a center mount, enabling to measure R and T simultaneously. The angle of incidence was varied between 7° and 77°, with a step of 10°. The spot size was around 6 $mm^2$ and the spectral step 1 nm. The electrical performances of the cells were evaluated by quantum efficiency measurements with an aperture area of 1 $mm^2$, and current–voltage (I–V) measurements under a calibrated AM1.5 solar simulator, a Wacom class AAA (spectral match, time instability, and nonuniformity of irradiance) with a collimating angle below 5°.

## III. Results and Discussion

The topography of the nanopattern was inspected just before and after front-side cell fabrication. The optical and electrical performances of the nanopatterned and reference (flat) cells were, then, measured and confronted to the modeled results.

### A. Topography: Deviation from ideality

We checked how the nanopattern of the cell differed from the ideal model, that is the hole dimensions, their defects, and the conformality of the coatings on the top. As identified in [21], the ideal 2-D pattern, which would maximize the integrated absorption and that was targeted experimentally, is a square lattice of circular holes and it has the following characteristic dimensions: period ~540 nm, diameter ~350 nm, and depth ~110 nm.

Inspection by SEM showed the lattice had a large spread of dimensions in diameter, with imperfect circular shapes and rough walls, while the pitch was as targeted (see Fig. 2). The hole depth, as measured by AFM, was smaller than targeted and between 85 and 95 nm. This deviation of depth from target can be attributed to the fact that this depth could not be monitored in-situ.

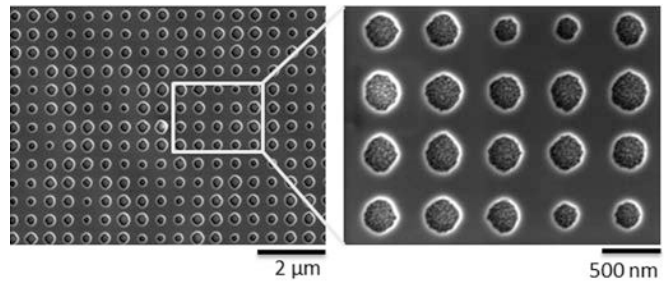

**Fig. 2. 2D-patterned silicon surface with a pitch of ~600 nm after mask removal and before a-Si:H and ITO deposition, with its irregular hole shape and rough walls (SEM).**

Conformality of the layers deposited on the top of the pattern was observed by TEM. The a-Si:H layer, deposited by PECVD, was more conformal than the ITO layer, deposited by sputtering (see Fig. 3). The thickness of a-Si:H is reduced by 40% on the vertical walls, while that of ITO may locally be only 10% of the horizontal thickness. The ITO, although locally very thin, is continuous over the whole surface (see Table I). It is also thinner than expected. TEM cross sections also show that the silicon profiles are not perfectly square but slightly slanted, with rounded corners. The morphology of these two coatings, and especially that of ITO, could have a significant impact on the c-Si layer absorption and should be taken into account when modeled. Optically, nonconformal

ITO may be preferred to reduce parasitic absorption [24], but electrically it would not enable proper electron transport.

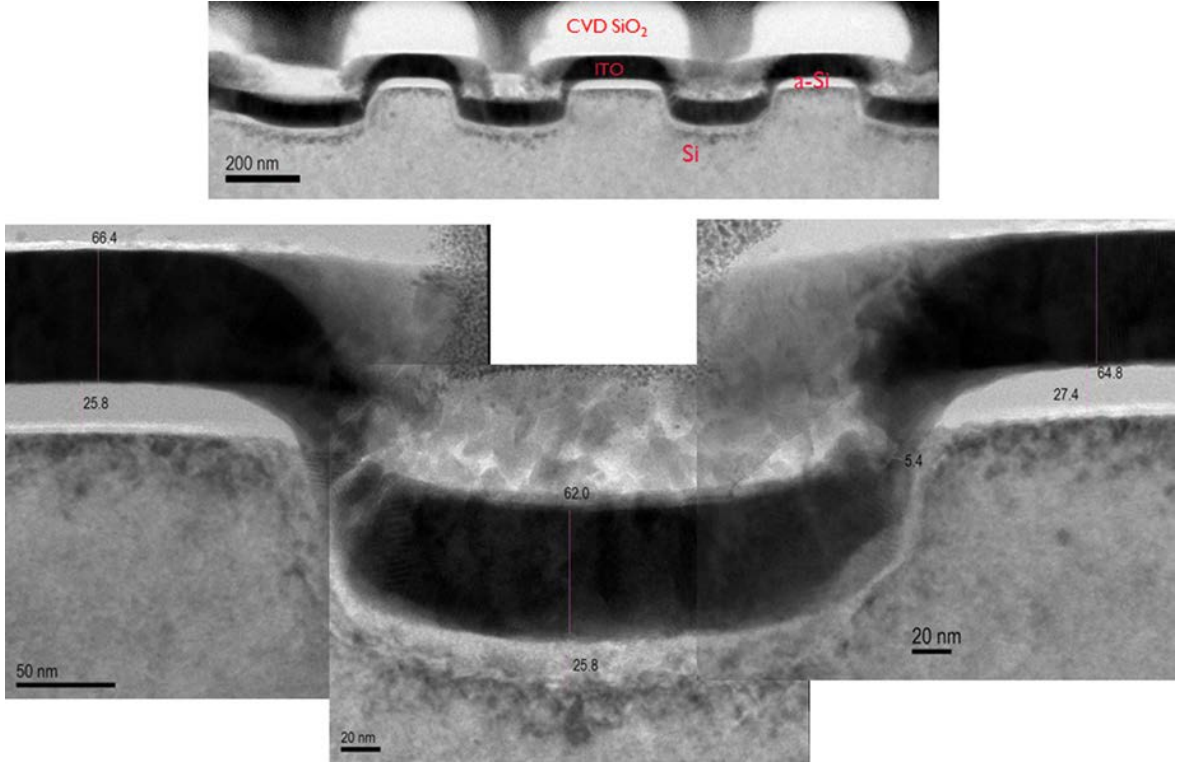

**Fig. 3. 2D-pattern covered by a-Si:H (light) and ITO (dark) in one sample location (TEM). The dark/bright contrast at the top of the crystalline Si is due to material deposition/roughness likely occurring during the process.**

**Table I: Variation of thickness of ITO and a-Si:H as measured by TEM in three different locations, showing the nonconformality of the ITO layer.**

| Layer thickness (nm) | Top       | Sidewalls  | Bottom    |
|----------------------|-----------|------------|-----------|
| ITO                  | 62.9-67.2 | 5.4-49.9   | 62.0-68.1 |
| a-Si:H               | 25.8-27.2 | 16.1       | 23.8-26.2 |

The experimental pattern that is obtained differed (in diameter, depth, shape, etc.) from the targeted pattern modeled in [21], and the simulations were, therefore, updated to this new configuration. Nonconformality of ITO had been expected and taken into account in [21]. As for the hole period, it is the most regular feature and it falls in the tolerance window of 0.55–0.65 μm that was previously found. The hole diameter shows a strong variation, but the tolerance window of the surface filling factor of 35%–60% is large and diameters are, therefore, also in an acceptable range. Profile depth, however, does not stay within the optimum window of 110–140 nm. The newly simulated stack is illustrated in Fig. 4.

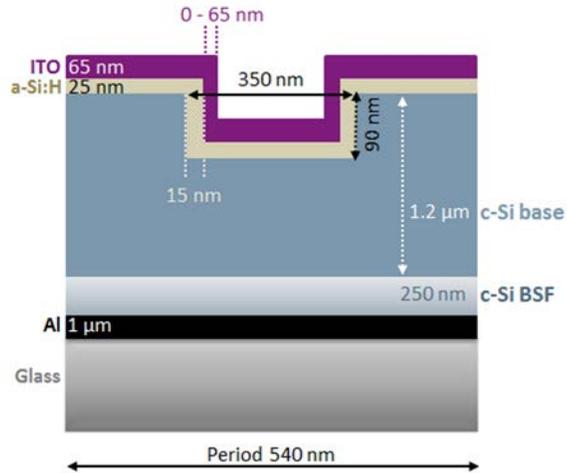

**Fig. 4. Solar cell stack, depicted over one period of the PC (540 nm), simulated by FDTD and matching the experimental stack, with a base thickness of 1200 nm, a hole depth of 90 nm, a BSF of 250 nm and ITO of 65 nm (not at scale). The ITO thickness along vertical walls is varied between 0 and 65 nm (from discontinuous to perfectly conformal). For a-Si:H, thickness on vertical walls is set at 15 nm and at 25 nm for horizontal surfaces.**

B. *Optical performances: Deviation from model and impact of two-dimensional photonic crystal*

The absorbance of the fabricated cells was first compared with that of their modeled counterpart, to check the validity of the model and to evaluate the effect of the imperfections observed in Section III-A. Then, the measured absorbance of the patterned sample was compared with that of the reference flat sample to assess the impact of the nanopattern.

A first confrontation between experiment and model is done by comparing the absorption spectra of flat samples (see Fig. 5a).

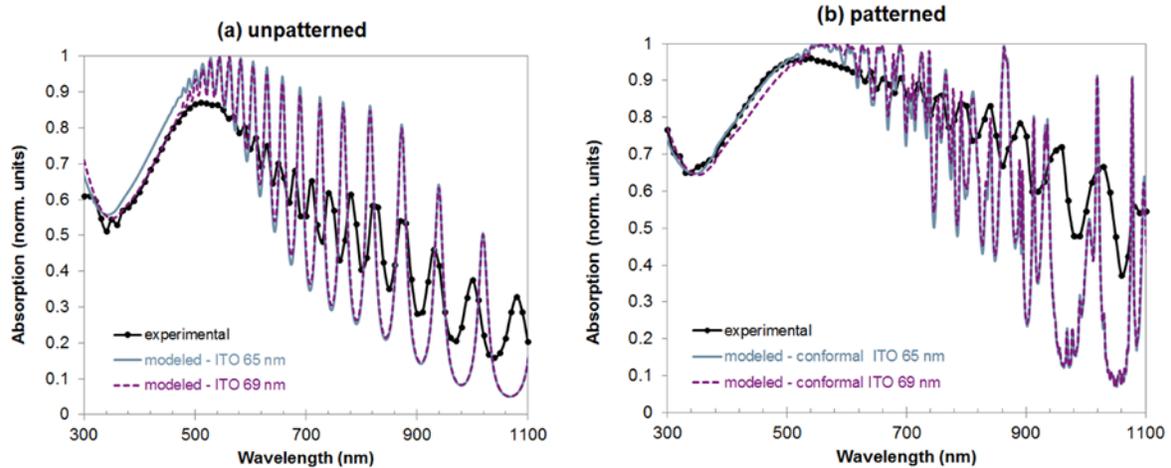

**Fig. 5.** (a) Absorption of the unpatterned and (b) patterned cell, measured (black circles) and modeled (no markers).

At short wavelength (i.e., up to ∼400 nm, with photons absorbed at the front side), the two curves do not perfectly match and show the need for a perfect tuning of the ITO thickness. The best match is obtained with 69 nm, which is close to the thickness measured by TEM on the patterned sample. As for the longer wavelengths (i.e., above∼500 nm, with photons reaching the rear side), peaks due to Fabry–Perot interferences appear. Their different free spectral ranges indicate that the refractive indexes used in the simulation do not match the actual ones for the Si/Al interface. Their lower measured intensity also suggests this interface may be different from the flat modeled interface. TEM images of the Si/Al interface suggest that this may be due to a reaction between Al and Si, witnessed by the presence of unexpected spikes of Al inside the c-Si (see Fig. 6). Their origin is still unclear but a possible cause may be the bonding process, which subjects the sample to 1000 V at 250 °C for 10 min and might have caused Al migration [22]. As for the spectral resolution, it can be expected to be reasonably sufficient not to be responsible for this discrepancy.

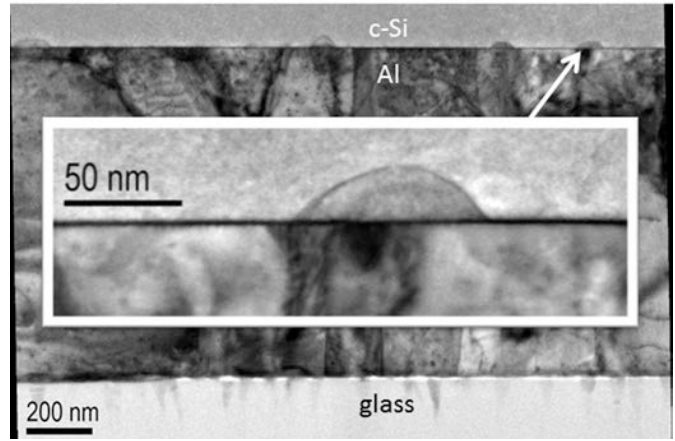

**Fig. 6.** c-Si/Al and Al/glass interface, showing unexpected spikes of Al inside the c-Si (TEM).

For the patterned cell [see Fig. 5(b)], peaks from the Bloch modes superpose to the peaks of the Fabry–Perot effect, and result in a more complex pattern. The loss of peak sharpness (broader and lower peaks) and the global increase of absorption may be attributed to the global dispersion of the hole diameters [25] as well as to the roughness and nonsteep walls, by increasing scattering and by a more progressive refractive index change. In addition, and in the long wavelength range only, the Si/Al interface can also induce loss of peak sharpness, as well as a shift between theoretical and experimental Fabry–Perot resonant wavelength [see Fig. 5(a)]. These two deviations from the ideal pattern may explain that experimentally the 2DPC improves absorption over a wider spectral range than theoretically.

Finally, the ITO layer may also be a source of deviation from the model. In fact, just as for the flat case, the thickness of ITO has a strong influence on the short wavelengths (cf., Fig. 5(b), 69 nm versus 65 nm). The thickness on the vertical walls was then tuned, to evaluate how strongly the nonconformality of the deposition influences absorption (see Fig. 7). At a glance, a thicker lateral ITO, on vertical walls, induced an increase in global absorption, especially in the short wavelength range. In addition, it can be noted that in the absence of lateral ITO, the interferences due to Fabry–Perot effects are more visible. A better fit of the model is obtained with thicker lateral ITO. This may seem surprising since the deposition as seen in Table I is not conformal, however, the actual profile is rounded (see Fig. 3), and therefore, different from the square approximation of the model.

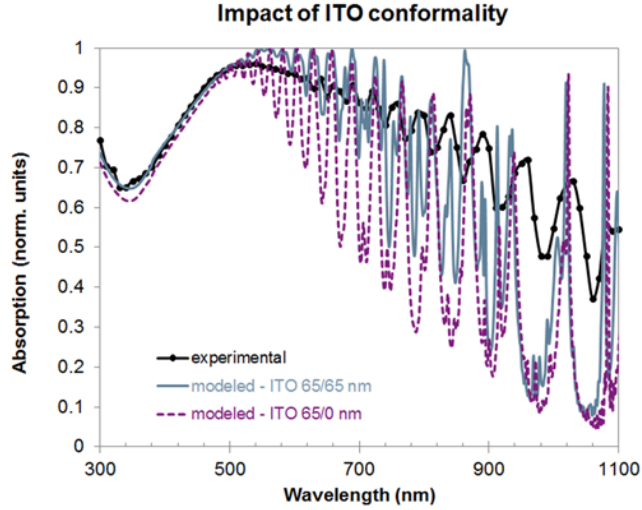

**Fig. 7.** Absorption of the 2DPC-patterned cell simulated for different thicknesses of ITO on the pattern vertical walls, from perfect conformality (65/65 nm) to non-continuity (65/0 nm).

Both theoretically and experimentally, inclusion of a 2DPC at the front side of the cell enhances cell absorption over the whole spectrum, but more particularly in the red part, for which it increases the photons lifetime, and thus, maximizes their absorption probability (see Fig. 8). One may, therefore, hope for a corresponding enhancement in the cell's external quantum efficiency (EQE). However, these absorption spectra (see Figs. 5, 7 and 8) are those of the complete solar cell stack. The useful absorption, corresponding to the absorption in the c-Si layer only, is only a part of these global absorption spectra. ITO and Al layers are known to parasitically absorb a significant part of the light, as calculated in [21] or [24]. We, therefore, extracted from the total cell absorption the absorption localized inside the various layers composing the cell and isolated that of the active material, the c-Si base (see Fig. 9). Contribution from the BSF is disregarded, since it is too highly doped for charges to be collected.

An insight into the short-circuit current density (Jsc) that may be expected from the cell is, then, given by the percentage of incident photons that are absorbed in the c-Si layer. Generally, the optical efficiency η can be defined in each layer as the ratio between the incident photon flux $\phi_i$ and the absorbed photon flux $\phi_a$ in the corresponding layer and can be expressed as

$$\eta = \frac{\phi_a}{\phi_i} = \frac{\int \frac{\lambda}{hc} S(\lambda) A(\lambda) d\lambda}{\int \frac{\lambda}{hc} S(\lambda) d\lambda}$$

with λ the wavelength, h the Planck constant, c the speed of light, S(λ) the normalized solar spectrum AM1.5G, and A(λ) the absorption coefficient in the corresponding layer. In our study, the integral is done from 300 to 1100 nm (corresponding to the bandgap of crystalline silicon). Finally, if we assume that each photogenerated carrier in the active c-Si layer is electrically collected (i.e., perfect collection efficiency), then we have $Jsc = e\phi$ with e the charge of one electron, Jsc the maximal achievable photocurrent for the absorbed photon flux $\phi$ in the c-Si layer. It can be noticed that a complete sunlight absorption would lead to a Jsc of 43.5 mA/cm².

The maximal theoretical achievable Jsc for various cell configurations, in function of ITO conformality, are summarized in Table II.

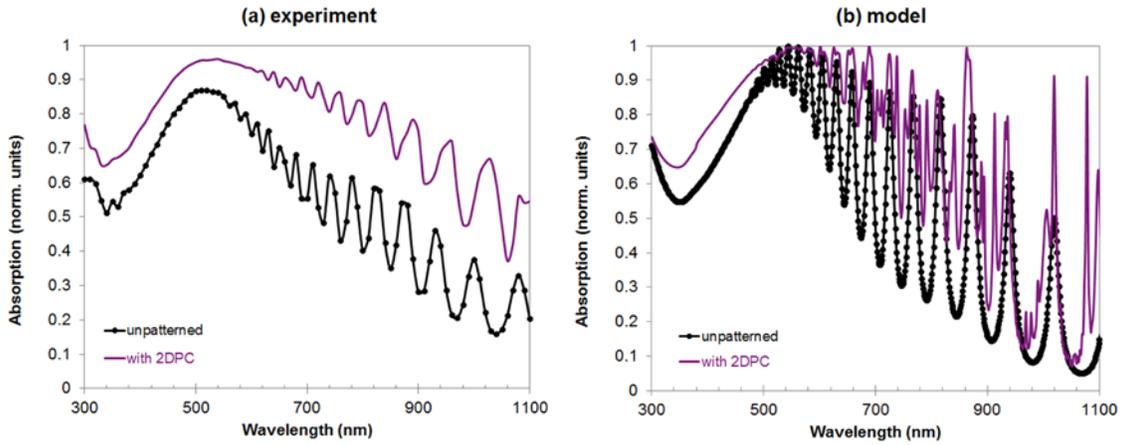

**Fig. 8. Measured (a) and simulated (b) global absorption improved by inclusion of a 2DPC in the solar cell stack. The unpatterned cell is (black circles) modeled with 69 nm ITO while the patterned cell (purple) is with 65/65 nm.**

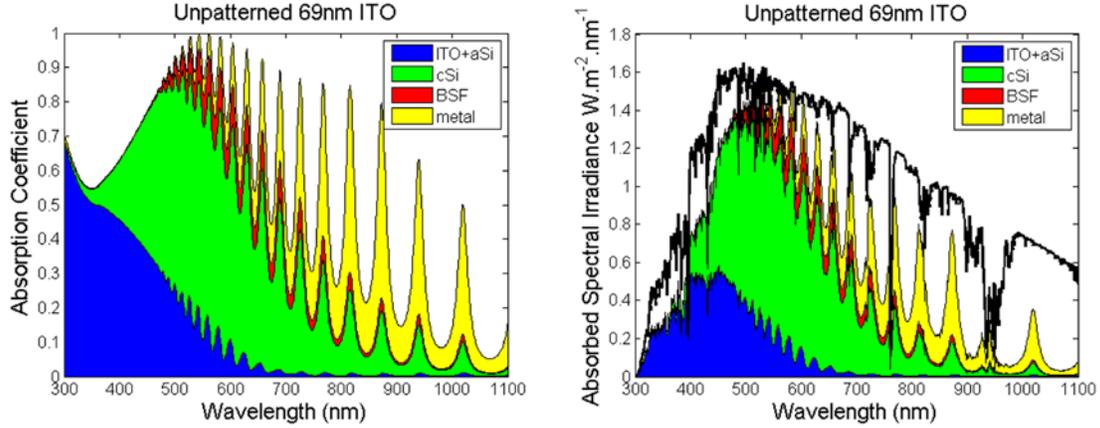

**Fig. 9. Modeled localized absorption spectra of the unpatterned cell in terms of absorption coefficient (left) and in terms of number of photons (right), isolating the contributions of each layer, from top to bottom, showing that ITO and a-Si:H (blue) are strong parasitic absorbers in the short wavelength range. The black curve corresponds to the AM1.5G spectrum.**

**Table II: Theoretical optical absorption efficiency in the whole cell and in each layer, and maximal achievable Jsc assuming perfect charge collection efficiency. Calculations are done for different cases of ITO conformality (thickness in nm). The (only) actual contributing layer, the base, benefits from conformal ITO.**

| Absorption efficiency (corresponding Jsc in mA/cm$^2$) / Cell | Global | Emitter (a-Si + ITO) | c-Si base | BSF | Al contact and reflector |
|---|---|---|---|---|---|
| Unpatterned | 54.6 | 8.7 | **27.8 (12.09)** | 4.1 | 14 |
| Patterned 65/65 (conformal ITO) | 71.4 | 10.9 | **35.8 (15.60)** | 5.5 | 19.1 |
| Patterned 65/50 | 68.8 | 10.6 | **33.1 (15.07)** | 5.3 | 18.2 |
| Patterned 65/30 | 65.2 | 10.3 | **32.9 (14.32)** | 4.9 | 17 |
| Patterned 65/00 (non-continuous ITO) | 60.4 | 9.9 | **30.6 (13.34)** | 4.6 | 15.4 |

The "optical efficiency" η corresponding to the absorption in the c-Si only, amounts to 27.8% for the unpatterned cell, while it reaches 35.8% for 2DPC cell with 65/65 nm of ITO. For comparison, it amounts to 33.1%, 32.9%, and 30.6% for the cases with 65/50, 65/30, and 65/0 nm of ITO, respectively. This means that, if both samples show the same charge collection efficiency (provided this integration did not alter the cell's electrical properties), we may expect an enhancement of 20% of the cell's Jsc by integration of the 2DPC.

Concerning conformality of ITO, we notice that conformal ITO not only increases the global absorption but also the useful absorption inside the c-Si base. Parasitic absorption, therefore, does not seem to overcome the function of antireflective coating, a result which interestingly is in contradiction with the findings of Yahaya et al. [24] (a possible reason for this contradiction could be the absence of a-Si:H in the cell stack of the latter).

Finally, we checked the robustness of the nanopattern effect on absorption with respect to the angle of incidence. As presented in [21], the integrated absorbance of the nanopatterned solar cell stays reasonably constant beyond 57° of incidence (see Fig. 10). We may notice that the absolute value of absorbance is lower than in [21], but that can be explained by the presence of metal fingers, that reflected part of the light because of the large spot size. It should also be noted that this robustness is an asset of nanopatterning as compared with traditional pyramid texturing, which performance drops more quickly at higher angles, i.e., cloudy illumination conditions.

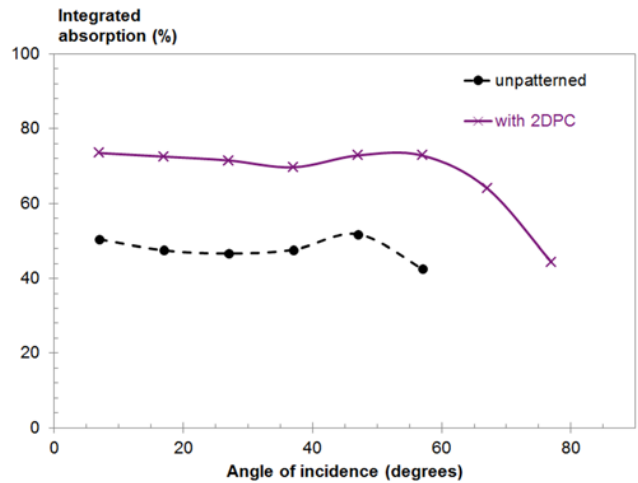

**Figure 10: Integrated absorption of the unpatterned (black circles) and nanopatterned (purple crosses) solar cells under various angles of incidence of light, showing the nanopattern stays efficient until a high angle.**

### C. *Solar cell performances: Photon collection vs. charge collection*

Illuminated current–voltage (I–V) curves show that thanks to the significant boost in light absorption, the short-circuit current is correspondingly improved. However, the overall cell efficiency dropped with the nanopattern. The cell's quantum efficiency suggests that the pattern induced a strong and deep surface degradation, leading to stronger surface recombination.

By inclusion of the 2DPC, the cell's Jsc rose by about 20%, which is in the range of what could be expected from the ideal model case. However, the open-circuit voltage (Voc) and the fill-factor (FF) dropped significantly (see Table III) pulling the efficiency down more than 20%. The shape of the curve (see Fig. 11) shows that both shunt and series resistances increased with the nanopattern (with an inflection in the curve suggesting the loss of ohmic contact [26]).We can suspect that the thinning of TCO and a-Si:H layers on the vertical walls may be responsible for this loss in FF. Patterns with less steep walls seem in fact less affected by this degradation [18]. It should be noted that the reference flat sample presents very good performances for this type of the cell, perfectly similar to what was obtained with this type of structure in the past [27]. We should also note that the measured Jsc values are larger than (or very close to) the maximum theoretical values discussed in the previous section. A first source of difference can be attributed to a different condition of irradiance, which is also illustrated in the difference between the solar simulator and the spectral response (in the solar simulator, the complete cell area is exposed to sunlight, while in the spectral response only a small area is exposed and, furthermore, to monochromated light). In fact, integration of the measured cell's spectra gives 12.08 and 14.35 mA/cm$^2$, respectively, which are close to the modeled values of 12.1 and 15.6. Other possible sources of difference could be the differences in refractive index of ITO or a-Si:H (which properties may strongly vary), the model integration limit of 1100 nm, or a possible contribution from the BSF. Finally, when comparing model and measurement, one should bear in mind that the AM1.5G spectrum used in the model is not perfectly simulated by the experimental light sources. Particularly, the sunlight divergence is lower than that of the simulator (0.5° versus 5°). The coherence area of the sunlight is thus about ten times larger, so closer to the plane wave used for the simulations. Experimental spectra in actual sunlight could, thus, be closer to the simulated ones, since the various peaks in these spectra result from interference phenomena. Similarly, the difference between the Jsc from spectral response and Jsc from I–V simulator might also be caused by the fact that the former was integrated over the AM1.5G spectrum, and not on the spectrum of the I–V simulator.

**Table III: Performances of an unpatterned reference and of the solar cell integrating a 2D photonic crystal. The 2DPC improved the Jsc but decreased the global cell efficiency. The Jsc for the modeled cells is the maximum achievable current density, assuming perfect charge collection efficiency.**

| Cell | Jsc [mA/cm$^2$] | Max. *modeled* Jsc [mA/cm$^2$] | Voc [mV] | FF [%] | Efficiency [%] |
|---|---|---|---|---|---|
| Flat - experimental | 12.8 | 12.1 | 548 | 66 | 4.6 |
| With 2DPC - experimental | 15.4 (+20%) | 15.6 (+29.0%) | 403 | 56 | 3.5 |

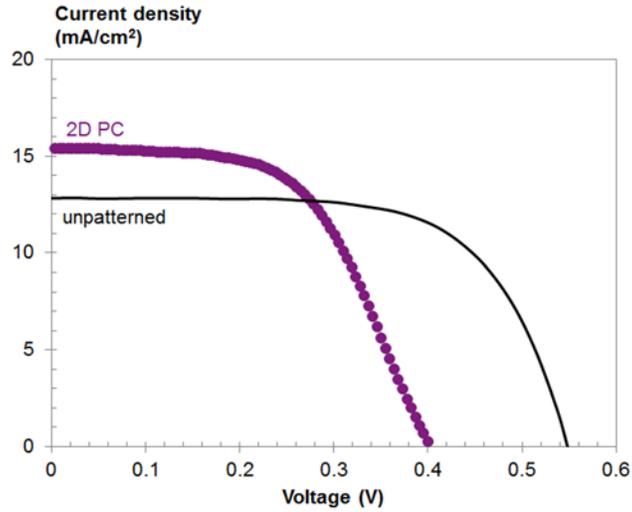

**Fig. 11. Integration of the 2DPC led to an improvement in Jsc but a degradation of Voc and FF.**

The cell's spectral response gives insight into the origin of the Voc degradation (see Fig. 12). One may guess that the enlargement of the front surface area must be responsible, at least partly, for a minority charge-carrier lifetime drop. However, there seem to be other sources of recombination. The internal quantum efficiency drops with the 2DPC until 600 nm that is a wavelength from which most photons reach the rear-side of the cell. The impact seems stronger on the shorter wavelengths, i.e., for wavelengths where photons are absorbed at the front side, where the pattern is located. However, the bulk of the cell is also visibly affected, confirming previous observations that plasma etching may damage the material up to a depth of ~1 µm [18], [28], which in our extreme case of an ultrathin cell corresponds to the bulk. The passivating performance of intrinsic a-Si:H as used in the current cell process is, therefore, not sufficient to compensate for the damage due to patterning. For long wavelengths, the EQE rises above the one of the reference cell, illustrating that the 20% increase in Jsc can be attributed to the red photons that the 2DPC can efficiently trap. However, at the front surface, the gain in reflectance from the 2DPC does not compensate for the loss, which leads to a lower EQE after patterning over short wavelengths. We may also note that the reflectance curve at the short wavelengths is strongly affected by absorption from the ITO and a-Si:H coatings, which do not contribute to the Jsc.

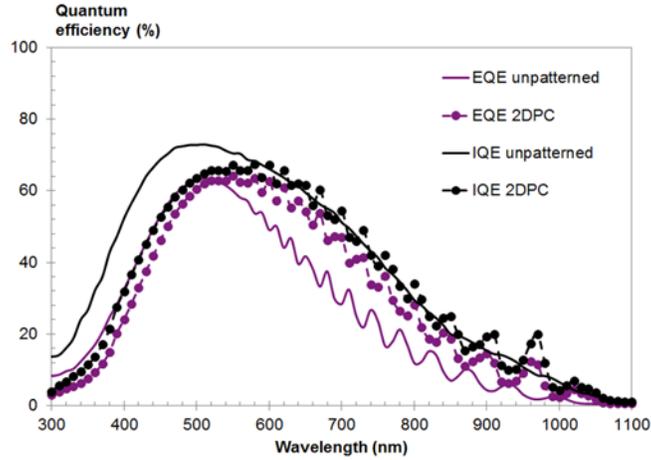

**Fig. 12. External and internal quantum efficiencies of unpatterned (plain) and patterned (circles) cells, with improved red photons absorption.**

## IV.   Conclusion

As expected and hoped for, integrating a 2DPC into a 1-μm-thick c-Si cell leads to a significant boost in the cell's Jsc, by a more efficient capture of red photons and by an improved antireflection effect over wide angles of incidence. However, this did not result in an overall cell efficiency boost, for the charge collection efficiency of the cell dropped. In fact, the minority charge-carriers recombination velocities increased both at the device front surface and, out of its extreme thinness, also in the bulk of the active layer, which leads to a drop in Voc. Besides surface quality, the nanotexture also impacted the quality of the TCO deposited on it, and thus, the FF. Localized absorption spectra confirmed that significant amounts of photons are lost by parasitic absorption from the TCO and the rear contact. We can also note that experimental deviations from ideality, such as the spread in structure dimensions, wall roughness, or variations in coating thicknesses, as experienced here, are to be taken into account in models, for they can significantly impact the optical behavior. Nevertheless, despite these deviations, we observe a significant increase in Jsc, confirming that the nanopattern concept developed here is quite robust to experimental uncertainties.

The impact of nanopatterns on thin-film c-Si cells, and especially on their surface recombination, demonstrates that combined optical and electrical simulations are needed for effective optimization of these structures. In fact, optically optimized surface morphologies (e.g., with high aspect ratios, like nanowires) may be very far from electrically optimized morphologies, and challenging for keeping charge collection efficient. For further improvement, our results indicate that efforts could be oriented toward three different directions. First, front-side passivation with excellent conformality, such as AlOx by atomic layer deposition, and means to protect surface from etch damage will have to be developed. Second, parasitic absorption at the rear side should be decreased, possibly by implementing local rear contacts or an interdigitated contact scheme (with emitter either at front or rear). Finally, parasitic absorption by the TCO at the front side may benefit from a newly optimized nanopattern taking into account evolution of its electrical properties.

## *Acknowledgments*

The authors would like to thank A. Herman, from the Université de Namur, for the angle-resolved absorption measurements, and their colleagues at imec, K. Van Nieuwenhuysen for the epitaxial deposition, and J. Geypen and H. Bender for the TEM inspection. This work was supported in part by the FP7 EC project PhotoNVoltaics under Grant 309127.